\documentclass[12pt,a4paper]{article}

\usepackage{times}
\usepackage{graphicx}
\usepackage{amssymb,amsmath}
\usepackage{mathrsfs}
\usepackage{fancybox}
\usepackage[latin1]{inputenc}
\usepackage{pst-all}
\usepackage{epsfig}
\usepackage{rotating}
\usepackage{subfig}
 
\newcommand{\be}{\begin{equation}}
\newcommand{\ee}{\end{equation}}
\newcommand{\bea}{\begin{eqnarray}}
\newcommand{\eea}{\end{eqnarray}}

\newcommand{\bdm}{\begin{displaymath}}
\newcommand{\edm}{\end{displaymath}}
\newcommand{\D}{{\rm d}}

\begin{document}

\begin{titlepage}

\noindent
\begin{center}
\vspace*{1cm}

{\large\bf SPACE AND TIME 62 YEARS AFTER THE BERNE
  CONFERENCE}\footnote{To appear in: {\em Thinking About Space and
    Time}, edited by C.~Beisbart, T.~Sauer, and C.~W\"uthrich
  (Birkh\"auser, Basel, 2020).}

\vskip 1cm

{\bf Claus Kiefer} 
\vskip 0.4cm
Institute for Theoretical Physics, University of Cologne
\vspace{1cm}

\begin{abstract}

In 1955, an international conference took place in Berne at which the 
state of relativity theory and its possible generalizations were
presented and critically discussed. I review the most important
contributions to that conference and put them into the perspective of
today's knowledge about the nature of space and time.
  
\end{abstract}

\end{center}

\end{titlepage}


\section{Historical Context}


 From July 11 to July 16, 1955, a conference took place in Berne
 celebrating the fiftieth anniversary
 of the special theory of relativity.
This was perhaps the first international conference
devoted to an overview of relativity theory, its ramifications, and
applications. The main goal of that conference was neither historical
nor was it restricted to special relativity; in fact, most of the
topics deal with general relativity and its generalizations, both in
classical and quantum directions, along with cosmology and with mathematical
structures. The list of participants contains an impressive number of
famous figures together with a selection of young
scientists.\footnote{One of the younger participants, Walter Gilbert,
  was awarded the 1980 Nobel Prize in chemistry.}
The Proceedings of that conference were published in 1956 and contain
most of the presented talks together with a record of the discussions
(Mercier and Kervaire~1956). In my contribution, I will heavily rely
on these Proceedings, the title page of which is displayed in Fig.~1. 

\begin{figure}[h]
  \includegraphics[width=0.6\textwidth]{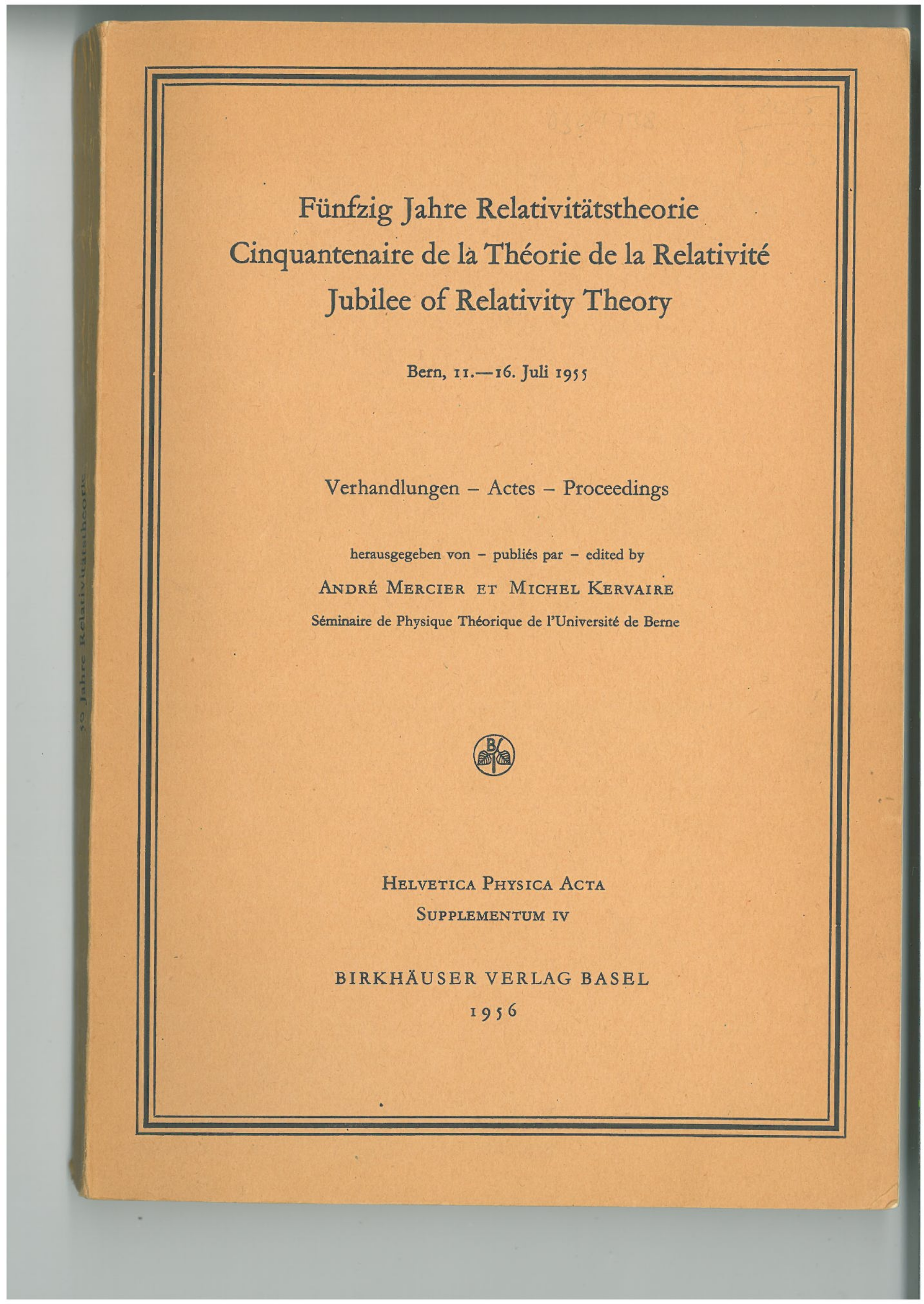}  
\caption[]{Title page from the Proceedings of the Bern
  Conference~1955.
\copyright\ Birkh\"auser Verlag, Basel. Photo by the author.}
\end{figure}

The Berne Conference was later known as the GR0 Conference, where the
numbering refers to the series of conferences organized by The
International Society on General Relativity and
Gravitation (GRG).\footnote{See isgrg.org} This society was founded at the
GR6 conference, which took place in Copenhagen in 1971, and grew out
of the International Committee on General Relativity and Gravitation,
which was responsible for the earlier conferences. Several of the GRG
Presidents were, in fact, participants of the Berne Conference, among
them Christian M\o ller, Nathan Rosen, Peter Bergmann, and Yvonne
Choquet-Bruhat. 
 
The conference was organized by the Seminar for Theoretical Physics
of the University of Bern, located at the Institute for Exact Sciences
at Sidlerstrasse~5. The President of the Organization Committee was
Wolfgang Pauli (Z\"urich), the Scientific Secretary Andr\'e Mercier
from Berne. The lectures themselves took place in the lecture hall of
the Natural 
History Museum. Figure~2 shows the building at Sidlerstrasse
at the time of the conference. It was there that Einstein delivered
his lectures as a {\em privatdozent} in Berne. These were the lectures
on molecular theory of heat ({\em Molekulare Theorie der W\"arme}) in
the summer term 1908 (with three attendants, which were his friends),
and on the theory of radiation in the winter term 1908/09 (with four
attendants, even including a student), see F\"olsing (1994, p.~274). 

\begin{figure}[h]
   \includegraphics[width=0.7\textwidth]
     {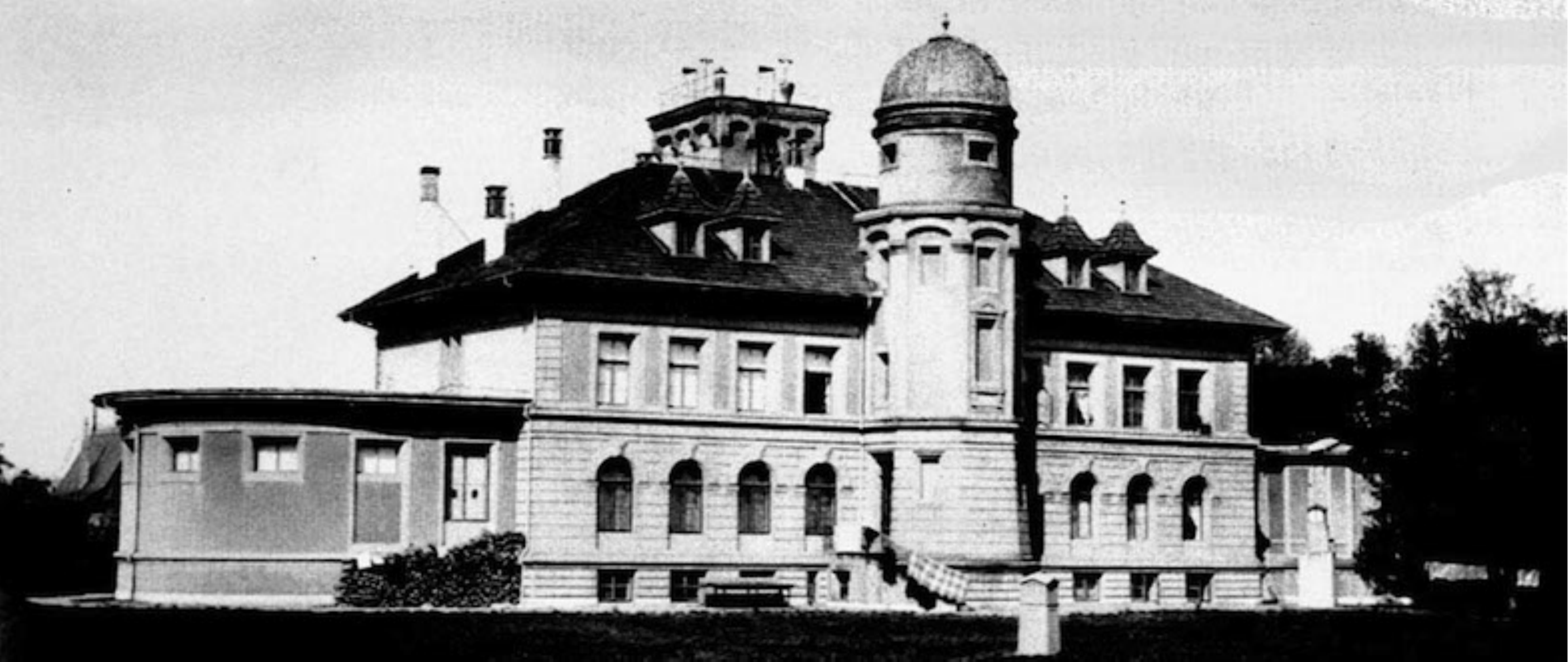} 
\caption[]{Contemporary look of the building at
  Sidlerstrasse~5. \\ \copyright\ Universit\"at Bern.} 
\end{figure}

 Between 1959 and 1963, the new building shown in Fig.~3 was erected,
 designed by the architect couple Hans and Gret Reinhard. It clearly
 demonstrates the new spirit of the 
 day.\footnote{See unibe.ch/university/portrait/history/}

\begin{figure}[h]
   \includegraphics[width=0.7\textwidth]
     {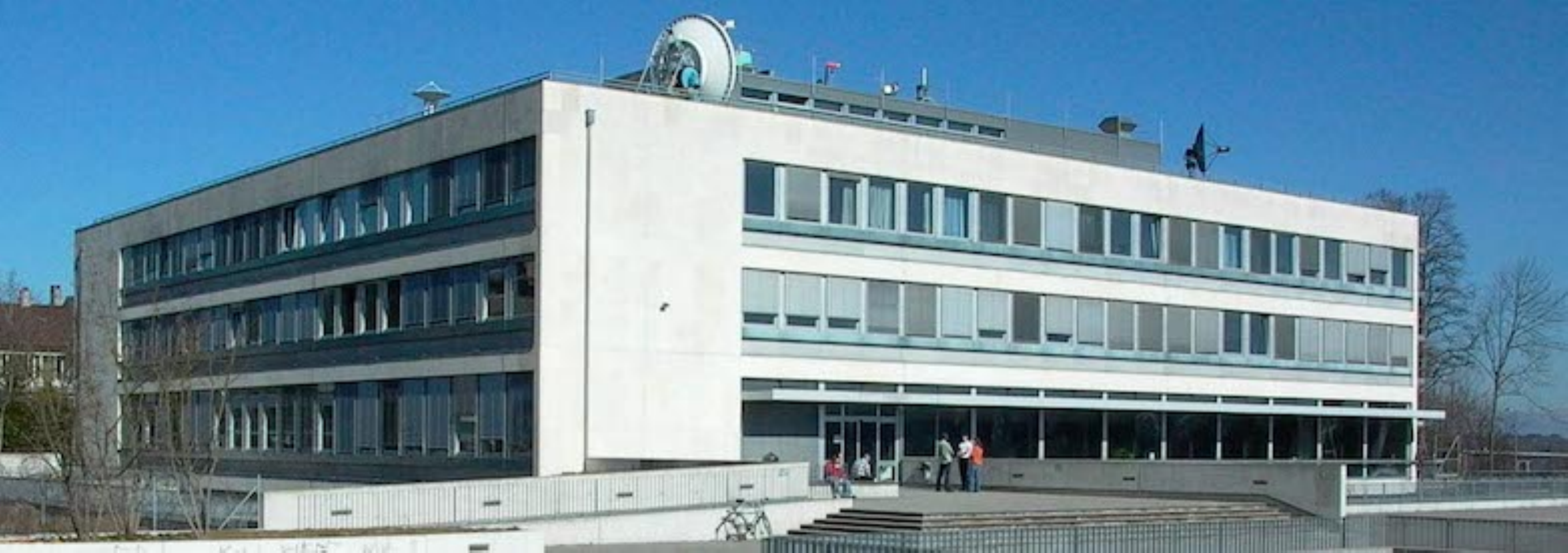}  
\caption[]{Present look of the building at Sidlerstrasse~5.
\copyright\ Universit\"at Bern.}
\end{figure}

In his foreword to the Proceedings, the secretary Andr\'e Mercier made
the following interesting remarks on the theory of relativity:

\begin{quote}
The theory of relativity marks all in all one term: it is the
achievement of a physics of cartesian spirit that gives an account of
the phenomena by figures and by motions \ldots
One hears today the saying that we live in the atomic era. Should we
not also speak of the relativistic era?\footnote{This is my
translation from the original French which reads:
``La Th\'eorie de la Relativit\'e marque en somme un terme: elle est
l'ach\`evement d'une physique d'esprit cart\'esien rendant compte des
ph\'enom\`enes par figures et par mouvements. \ldots 
On entend dire aujourd'hui que nous vivons \`a l'\`ere atomique. Ne
pourrait-on aussi bien parler de l'\`ere relativiste?'' (Mercier and
Kervaire (1956), p.~19)}
\end{quote}

This, of course, alludes to the fact that the theory of relativity was
not very popular in the 1950s, apparently overshadowed by quantum theory and its
applications to atomic and nuclear physics. 

Of interest is also the welcome speech of a local politician, a
certain Dr. V.~Moine, {\em Directeur de l'instruction publique du
  Canton de Berne}. He reflects the philosophical spirit of the
conference's location:

\begin{quote}
This city, enclosed like a jewel by the crown of the river Aare, with
its military and political past which made it the head of the old
Switzerland, practical and empirical as a farmer's wife, has always valued
the positive and immediate higher than the theoretical. Its symmetric
streets, its order, the equilibrium which is brought out by its
buildings, the traditional caution of its laws make it more a city of
Aristotle than of Platon. \ldots\footnote{This is my
translation from the original French which reads:
``Cette ville, enserrée comme un joyau dans la couronne de l'Aar, au
passé militaire et politique qui fit d'elle la tête de la vieille
Suisse, pratique et empirique comme une paysanne, a toujours plus
apprécié le positif et l'immédiat que le théorique. Ses rues
symétriques, son ordonnance, l'equilibre qui se dégage de ses
édifices, la prudence traditionelle de ses lois en font plus une cité
d'\textsc{Aristote} que de \textsc{Platon}. \ldots'' (Mercier and
Kervaire (1956), p.~25)}
\end{quote}

(The building shown in Fig.~3 perhaps also gives testimony of this
Aristotelian attitude.)
This practical spirit is also seen in the organization of the
conference. Three languages (English, French, German) are used
interchangeably, and in 
the discussions they are often mixed in an interesting way;
an example is the discussion after Bergmann's talk with its m\'elange of
the three languages (Mercier and Kervaire (1956), pp.~95/96). 

Albert Einstein had been invited to this conference, but he died on
April 18, 1955, three months before the Berne Conference.
Anyway, he had not envisaged to attend the meeting. In his reply to a letter
of invitation by Louis Kollros,\footnote{Louis Kollros (1878--1959) was
  a Swiss mathematician; from 1909 to 1948 he was professor at ETH
  Z\"urich.} Einstein had written:

\begin{quote}
We two are no spring chickens anymore! As for me, I cannot think about a
participation. \ldots\footnote{This is my translation from the original
  German which reads: ``Wir sind beide keine Jünglinge mehr! Was
  mich betrifft, so kann ich nicht an eine Beteiligung
  denken. \ldots''
(Mercier and Kervaire (1956), p.~271)} 
\end{quote}

It is left entirely to our imagination to figure out what would have
happened if Einstein had been able to attend the Berne meeting.


\section{Classical General Relativity and Beyond}

The year 1955 marked the 40th anniversary of Einstein's general theory
of relativity. Since it was difficult in those years to test the
theory empirically beyond the classic tests (redshift, light
deflection, perihelion motion), much attention was focused on
theoretical and mathematical developments. This concerned, in
particular, the structure of the Einstein field equations, notably the
initial value problem and the problem of motion. As for the former,
two of the main figures, Andr\'e Lichnerowicz and Yvonne
Choquet-Bruhat (at that time Four\`es-Bruhat) were present at the
meeting. As for the latter, Leopold Infeld was the main figure who was
present.  

The well-posedness of the initial value problem (Cauchy problem) is of great
importance. Only if there exist initial data, that is, data on a
three-dimensional hypersurface that determine the evolution according
to the Einstein equations uniquely, can one use the theory to
predict physical processes, for example, the emission of gravitational
waves from coalescing compact objects. Today, well-posedness is
generally granted as established, see, for example, Isenberg (2014).
It is a key ingredient in numerical relativity. 

By the time of the Berne Conference, a first theorem on the initial value
problem had already been proven by Choquet-Bruhat in 1952. A more general
theorem was proven in a 1969 paper of Choquet-Bruhat and Robert
Geroch, see Isenberg (2014) and Chru\'{s}ciel and Friedrich
(2004) for a detailed discussion and references.\footnote{The
  subtitle of the volume 
Chru\'{s}ciel and Friedrich (2004) in fact reads ``50 Years of the
Cauchy Problem in General Relativity''.} The theorem proven by
Choquet-Bruhat and Geroch can be stated as
follows (see Isenberg (2014), p.~307):
\vskip 2mm

\noindent
{\bf Theorem}: For any smooth set of initial data $(h_{ab},K_{cd})$,
where $h_{ab}$ 
is the three-metric and $K_{cd}$ is the extrinsic curvature (second
fundamental form), on a specified
three-manifold which satisfies the vacuum constraint equations, there
exists a unique (up to diffeomorphism) maximal globally hyperbolic
development. 
\vskip 2mm

Further developments are discussed in the above cited references. One
concerns the extension to the non-vacuum case: the theorem also holds,
for example, for the physically relevant case of the Einstein--Maxwell
theory. On the mathematical side, it has been shown that the required
degree of regularity can be weakened. Other developments concern the
stability of Minkowski spacetime under long-time evolution, the
stability of de~Sitter space, and investigations on the cosmic
censorship conjecture. The latter conjecture -- in its weak form
stating that the singularities arising from gravitational collapse
cannot influence future null infinity -- was formulated by Roger
Penrose in 1969. Most of these later investigations made heavy use of
the global methods (Penrose--Carter diagrammes) developed in the
1960s, which were unavailable at the time of the Bern Conference.

Currently, there is much interest in  classical generalizations of general
relativity; concrete examples are the $f(R)$ theories, where $R$ is
the Ricci scalar. 
Whether those theories also enjoy a well-posed initial
value problem is far from clear. It is thus too early to
make statements about the range of validity for those theories. It is
imaginable that they can only be applied in more restricted situations
and not, for example, to the non-perturbative treatment of
gravitational wave emission. 

The subject of gravitational waves, which after their first direct detection in
2015 is of central importance
today, received little attention in 1955. Nathan Rosen, in his talk,
basically reviewed his work with Einstein of 1937 in which they had
expressed doubts about the existence of gravitational waves in the
full non-linear theory. According to them, the plane wave solutions of
the linearized theory do not correspond to any exact solution of the
full theory. Today we know, of course, that they were in error and
that gravitational waves indeed exist. 

The experimental situation with general relativity was not in a good
shape at the time of the conference. Still,
the state of the art of the two classic tests concerning
gravitational redshift and light deflection (as well as the state of
cosmology, see below) was addressed. The gravitational redshift (time
dilation) is, 
for a constant field with gravitational acceleration $g$, given by the
standard formula

\be
\frac{\Delta\nu}{\nu}=\frac{gh}{c^2}.
\ee

In his contribution, Robert Trumpler reported on recent (from 1954)
observations in the spectral lines of the white dwarf 40~Eridani~B and
gave a list with the redshifts measured for 18 other stars. The
historic experiments by Pound and Rebka, determining the redshift in a
laboratory experiment using the newly discovered M\"ossbauer effect,
were still four years ahead. Those new types of experiments are of much
higher accuracy than the stellar observations which have thus lost
their significance. Today, the gravitational redshift effect is part
of everyday life, for example through the use of
the Global Positioning System (GPS). For a detailed discussion of the
current experimental situation, see Will (2014); frequency shifts have
been measured over a height of 1/3 of a metre. 

The second classic test discussed at Berne was light deflection. For a
grazing ray near the Sun, the deflection angle is given by
\be
\delta=\frac{4GM}{Rc^2}\equiv \frac{2R_{\rm S}}{R},
\ee
where $R$ is the solar radius, $M$ the solar mass, and $R_{\rm S}$ is the Schwarzschild radius. 
It is convenient to parametrize this effect by a post-Newtonian
parameter $\gamma$, which assumes the value $\gamma=1$ in general
relativity,
\be
\delta\approx \frac{1+\gamma}{2}1.''7505.
\ee 
The first observations were the famous ones performed at Sobral and in
Principe on the occasion of the solar eclipse on May 29, 1919. The
accuracy there was about 30~percent. In his talk at Berne, Trumpler reported
about results from other eclipses, those of 1922, 1929, 1947, and
1952. The accuracies there were not much better than in 1919. Today,
light deflection has been confirmed to an accuracy of 0.01~percent
(Will~2014). This is mainly due to the development of
very-long-baseline radio interferometry (VLBI).
Still, observations during the total solar eclipse of August 21, 2017
in the USA have led to a value for the light deflection of $1.7512$
arcsec, with an uncertainty of only 3\% (Bruns~2018).

As is evident from
Will (2014), the progress in experimentally testing general relativity
since 1955 has been tremendous, and the status of the theory in this
regard is similar to elementary particle physics.

There have been many developments since that no one could have
imagined in 1955. These concern, in particular, the field of
relativistic astrophysics, which more or less started in 1963 with the
discovery of the first quasar 3C 273 by Maarten Schmidt. For the study
of active galaxies, neutron stars, and black holes, general relativity
has proven indispensable. The very concept of a black hole was not
understood in 1955 and played no role at the conference. Today, we can
insight the coalescence of black holes and neutron stars by investigating
the gravitational waves they emit. A single black hole can be studied
by its influence on the surroundings; a prominent example is the
supermassive black hole in the centre of our Milky Way (Eckart {\em et
  al.}~2017). 

At the conference, some interest was also devoted to classical
generalizations of Einstein's theory. In the last few decades of his
life, Einstein himself was very much concerned with attempts to constructing a
unified field theory of gravity and electromagnetism. One might thus
have expected that those attempts (and similar ones by Schr\"odinger
and others) met with great interest at Berne.
But this was not the case. Only Bruria Kaufman, Einstein's last
collaborator, gave a main talk on the mathematical structure of the
non-symmetric field theory, in which the Christoffel symbols
$\Gamma^s_{ik}$ are not required to be symmetric. The discussion after
that talk contains only one mathematical comment from Marie-Antoinette
Tonnelat. 

The indifference towards Einstein's final attempts can well be understood. 
It had become obvious at least since the 1920s that quantum theory is
needed to describe the atomic and subatomic world. The strong and weak
interactions relevant for the microscopic regime are not taken into
account in Einstein's work. Most physicists thus suspected 
(rightly) that an essential part of the world was missing in
Einstein's attempts at a classical unification of gravity and
electromagnetism. This opinion is clearly expressed in a letter of
Pauli to Einstein from September 19, 1946:

\begin{quote}
My personal opinion still is \ldots that the classical field theory in
every form is a squeezed out lemon, out of which it is impossible to
get anything new!\footnote{This is my translation from the original
  German which reads: ``Meine pers\"onliche \"Uberzeugung ist nach wie
  vor \ldots, da\ss\ die klassische Feldtheorie in jeder Form eine
  v\"ollig ausgepre\ss te Zitrone ist, aus der unm\"oglich noch etwas
  Neues herauskommen kann!'' (von Meyenn (1993), p.~384)} 
\end{quote}

 From a modern point of view, a more promising idea for a generalization
was presented at the Berne meeting by Pascual Jordan. He gave an
example of a theory with a `varying gravitational constant'. Such a
theory can be represented as a
{\em scalar-tensor theory} of gravitation, in which a
scalar field $\phi$ is added to the gravitational sector
(see e.g. Fuji and Maeda (2003)). The action
for such `Jordan--Brans--Dicke theories'' (as they were called later
after the contributions by Brans and Dicke) reads
\be
S_{\rm JBD}=\int\D^4x\sqrt{-g}\left(\phi R-\frac{\omega}{\phi}\phi_{,\mu}
\phi_{,\nu}g^{\mu\nu}+{\mathcal L}_{\rm m}\right),
\ee
where $\omega$ is a new dimensionless parameter of the theory. Such
theories (and generalizations thereof) are of much interest today, for
example in connection with dark energy or the assumed inflationary
phase of the early universe.

At the Berne conference, Jordan's contribution was received with
scepticism. In his conference summary, Pauli `buried' (an expression
by Engelbert Sch\"ucking)\footnote{See Harvey (1999), p.~11.} Jordan's theory as follows:
 
\begin{quote}  
By the magic of his mathematical theorems, Mr. Jordan has
unfortunately prevented us from hearing something about his physical
reasons to assume a variation of the gravitational constant; this
would have surely interested all of us \ldots\footnote{This is my
  translation from the original German which reads:
``Nun hat uns leider Herr Jordan mit dem Zauber seiner mathematischen
Sätze verhindert, etwas darüber zu hören, was eigentlich seine
physikalischen Gründe sind, um eine Veränderung der
Gravitationskonstante anzunehmen; das hätte uns ja sicher alle sehr
interessiert \ldots''. (Mercier and Kervaire (1956), p.~265)}
\end{quote}

In our days, investigations into the variation of fundamental
constants find general acceptance. The main reason for this is the
expectation that a more fundamental theory than general relativity
arises from the implementation of quantum theory (see below). In some
of these theories, ``constants'' of Nature are described at high
energies by time- and space-dependent fields. Despite various
searches, however, no time or space variation of ``constants'' was
observed so far.


\section{Cosmology}

With the advent of Einstein's theory of general relativity, it was
for the first time possible to provide a consistent description of the
Universe as a whole. Assuming the cosmological principle, one arrives
at the `Robertson--Walker form' of the metric, from which the
`Friedmann--Lema\^{\i}tre equations' can be obtained from
the Einstein equations. Today, one often speaks of 
`Friedmann--Lema\^{\i}tre--Robertson--Walker' (FLRW) world models. On
the observational side, not much was known at the time of the Berne
Conference beyond Hubble's law and some crude age determinations.
Still, cosmology was an important topic at the conference, certainly
more important than one would have expected in retrospect. There were
major reviews 
by Walter Baade (Mount Wilson Observatory) from the observational and
by Howard Robertson (California Institute of Technology) from the
theoretical side.  
Baade did not deliver a manuscript to the Proceedings, so no
statements can be made about his contribution. Robertson has sent a
detailed manuscript that also contains a comparison of theory with
observation. 

It is not surprising that Robertson based his analysis on the
homogeneous and isotropic Robertson--Walker metric. But he included
the following important comment (p.~135 of the Proceedings):

\begin{quote}
It is to be emphasized that we have not {\em required} the real
universe to be one satisfying the uniformity conditions imposed above;
we are merely examining the nature of the idealized model of the real
world in which the obvious and all-important inhomogeneities are
ironed out. We are not imposing the uniformity as a `cosmological
principle' \ldots to which the real world must adhere.
\end{quote}

In his contribution, Robertson discusses the observational status from
a rather modern point of view. He presents a diagramme displaying the
age of the universe against the matter density, and he allows for any
value of curvature and cosmological constant $\Lambda$.  The empirical
value of the Hubble constant, $H_0$, at that time was given by
$180\ {\rm km}/{\rm s\ Mpc}$, much higher than today's
value.\footnote{The current value from the Planck Collaboration (2018)
  is $(67.4\pm 0.5)\ {\rm km}/{\rm s\ Mpc}$, a bit more than one third
  of the 1955 value. There is currently a tension between cosmological
  and non-cosmological measurements of $H_0$. With the Hubble Space
  Telescope and the Gaia parallax measurements one gets
$(73.52\pm 1.62)\ {\rm km}/{\rm s\ Mpc}$, see Riess {\em et al.}
(2018).} The discrepancy of the historic value with today's value lies
in the very crude distance measurements of the day, which have greatly
improved since then. 

Given the (too high) value of $H_0$ and conservative lower limits for
the age of the Universe, Robertson finds that ``\ldots we are forced
to reintroduce $\Lambda>0$ in order to save this time scale \ldots''.  
Today we know that $\Lambda$ (or its generalization in form of dark
energy) is positive, although for a different reason than 
the one given by Robertson: it is because the Universe is found to be
currently accelerating.  

In addition to Baade's and Robertson's overviews, various shorter
contributions on cosmology have been presented at the conference,
including talks by Max von Laue, Oskar Klein, and Otto
Heckmann. 
Heckmann, for example, presented a world model of Newtonian cosmology with
expansion and rotation, which he had developed together with Engelbert
Sch\"ucking. They had found that the introduction of rotation leads to a
model without initial (`big bang') singularity. One might therefore
wonder whether this can also happen in general relativity. This is,
however, not the case. As the singularity theorems proven in the 1960s
by Roger Penrose and Stephen Hawking show, singularities are
unavoidable given some general assumptions. But those theorems were
not available at the time of the Berne Conference. 

In 1955, the expansion of the Universe was not yet generally
accepted. Ten years before the discovery of the Cosmic Microwave
Background (CMB), it was still possible to seriously uphold the alternative
model of a steady-state universe. Two of the main proponents of that
model, Hermann Bondi and Fred Hoyle, were present in Berne and gave two
short contributions. Today, this is of historic interest only. 

Max Born, in his account of ``Physics and Relativity'',
describing personal remembrances of the years around 1905, remarks
that the importance of general relativity lies in the revolution which
it has produced in cosmology. This is a remark that can certainly be
appreciated today much more than in 1955.


\section{Quantum Theory and Gravity}

In the 1950s, quantum theory and its applications were at the centre
of physics research worldwide. This is not surprising. In the realm
of atoms, nuclei, and particles, plenty of new experimental results
were found, and quantum theory, in its mechanical as well as field
theoretical version, was believed to be the correct theory for their
description. It is for this reason that the relation between quantum
theory and relativity was discussed at length at the Berne Conference,
too. 

Eugene Wigner presented a major lecture on ``Relativistic invariance
of quan\-tum-mechanical equations''. He not only reviewed his important work on
the representations of the Poincar\'e group but also discussed its
extension to the de~Sitter group. This is very interesting from a
modern point of view, because current observations indicate that our
Universe is asymptotically approaching a de~Sitter phase. Wigner
emphasized that massive particles must then be characterized by the
statement that their Compton wavelength is much smaller than $c/H$,
where $H$ is the (constant) Hubble parameter of de~Sitter space. 
The relevance of de~Sitter space for the formulation of asymptotic
conditions is emphasized in Ashtekar {\em et al.} (2015).  

A major problem is the consistent unification of quantum theory
with gravity. This was open in 1955 and is still open today, in spite of
much progress that has happened since then. Only two major talks were devoted to this
problem, by Peter Bergmann and by Oskar Klein. Bergmann reviewed the
state of the canonical formalism, which can serve as the starting
point for the quantization of the gravitational field. This formalism
was pioneered by L\'eon Rosenfeld in Z\"urich in 1930 and later
developed in parallel by Bergmann and his group in Syracuse and by Paul
Dirac in Cambridge as well as by Arnowitt, Deser, and Misner
(ADM) in the United States.\footnote{See, for example, Kiefer (2012) for
  details.}  
Bergmann's talk was a bit dry in the sense that he restricted himself to
pure formalism and did not address physical applications. 

The real starting point of quantum gravity research is marked by a
conference two years later. It took place at Chapel Hill, North
Carolina, and was later known as the GR1 Conference. Many of the
proponents of quantum gravity were present, including John Wheeler and
Bryce DeWitt. Richard Feynman discussed there a gedanken experiment from
which he concluded the necessity for quantizing the gravitational
field, see Fig.~4.

\begin{figure}[h]
   \includegraphics[width=0.8\textwidth]{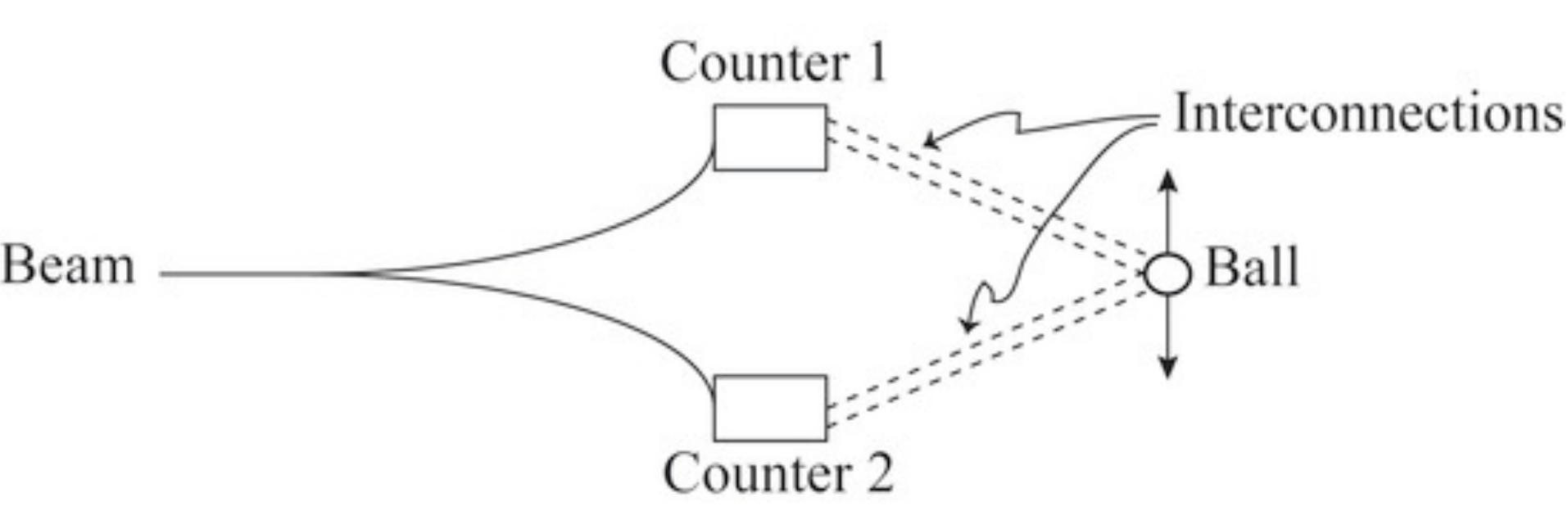} 
 \caption[]{Stern--Gerlach type of gedanken experiment, in which the
   detectors for spin up respective spin down are coupled to a
   macroscopic ball. If the particle has spin right, which corresponds
 to a superposition of spin up and down, the coupling leads to a
 superposition of the ball being moved up and down, leading to a
 superposition of the corresponding gravitational fields. Figure
 adapted from DeWitt and Rickles, p.~251, see DeWitt (1957).} 
\end{figure}

Feynman concludes:\footnote{See also the discussion in Feynman {\em et
    al.} (1995).}

\begin{quote}
\ldots if you believe in quantum mechanics up to any level then you
have to believe in gravitational quantization in order to describe
this experiment. \ldots It may turn out, since we've never done an
experiment at this level, that it's not possible \ldots that there is
something the matter with our quantum mechanics when we have too much
{\em action} in the system, or too much mass---or something. But that
is the only way I can see which would keep you from the necessity of
quantizing the gravitational field. It's a way that I don't want to
propose. \ldots (DeWitt and Rickles, p.~251-2, see DeWitt (1957))
\end{quote} 

 The Berne Conference was still far behind this level of physical
 discussion. But in his concluding speech, Wolfgang Pauli expressed very
 clearly the main difficulty in quantizing the gravitational field. 
He said:

\begin{quote}
This now leads to the border of knowledge, to the questions of the
quantization of the field; it seems that a certain agreement existed
in assuming that the mere application of conventional quantization
methods probably will not lead to the goal. \ldots

It seems to me \ldots that it is not so much the linearity
or non-linearity which forms the heart of the matter [the difficulty
of quantizing the gravitational field, C.K.], but the
very fact that here a more general group than the Lorentz group
is present.\footnote{This is my
translation from the original German which reads:
``Das führt nun hier an die Grenze des Wissens, an die Fragen der
Quantisierung des Feldes; es scheint, daß eine gewisse Übereinstimmung
darüber bestand, daß eine bloße Anwendung konventioneller
Quantisierungsmethoden wahrscheinlich nicht zum Ziele führen wird.
\ldots

Es scheint mir \ldots, da\ss\ nicht so sehr
die Linearit\"at oder Nichtlinearit\"at Kern der Sache ist, sondern
eben der Umstand, da\ss\ hier eine allgemeinere Gruppe
als die Lorentzgruppe vorhanden ist.'' (Mercier and Kervaire (1956), p.~266)}
\end{quote}

Standard ways of quantization assume the existence of a fixed
background, which usually is taken to be Minkowski space. In general
relativity, this background is absent -- spacetime is dynamical, and
the invariance is the diffeomorphism, not the Lorenz group. The
quantization of the metric (which represents spacetime) has thus to be
undertaken without any reference to Minkowski space with its Lorentz
group; this is what Pauli is alluding to. Modern approaches to quantum
gravity make use of this background independence (Kiefer~2012).

In this connection, it is interesting to quote a piece from a
letter that Pauli wrote to Schr\"odinger on the occasion of the
latter's 70th birthday. In this letter, which is from August 9, 1957,
he writes (von Meyenn~2011, p.~720):\footnote{I thank Norbert Straumann
  for drawing my attention on this and the following letter.}

\begin{quote}
Also our difference in age of 13 years will soon appear as
unessential, and one will count us as belonging to the same generation
of physicists: to those who have e.g. not succeeded in making a
synthesis of the mentioned subjects -- general relativity and quantum
theory -- and who thus have left behind unsolved problems as essential as
the atomistic nature of electricity (fine structure constant),
self-energy of the electron \ldots\footnote{This is my 
translation from the original German which reads: ``Auch unser
Altersunterschied von 13 Jahren wird bald als unwesentlich erscheinen,
und man wird uns zur selben Physiker-Generation z\"ahlen: zu derjenigen,
der z.B. eine Synthese der beiden genannten Themen -- allgemeine
Relativit\"atstheorie und Quantentheorie -- nicht gelungen ist und die
so wesentliche Probleme wie Atomistik der Elektrizit\"at
(Feinstrukturkonstante), Selbstenergie des Elektrons \ldots ungel\"ost
zur\"ucklie\ss .''}
\end{quote}

In his response from August 15, 1957, Schr\"odinger writes (von
Meyenn~2011, p.~722):

\begin{quote}
  You are, of course, right, that we belong to the same generation
  of physicists; I also agree with your characterization of it. But
  the posterity usually judges in a milder way, it characterizes an
  epoch by what it has achieved, much more rarely by what it has not
  completed.\footnote{This is my
    translation from the original German, which reads:
    ``Nat\"urlich hast Du recht, da\ss\ wir zu derselben
    Physikergeneration geh\"oren; auch dem, wie Du sie kennzeichnest,
    stimme ich bei. Nur pflegt die Nachwelt milder zu sein, sie pflegt
    eine Epoche zu charakterisieren nach dem, was sie geleistet hat,
    viel seltener nach dem, was sie nicht fertig gebracht hat.''}
  \end{quote}

Even today, the problem of quantum gravity remains unsolved.
The main approaches, more or less promising, are direct quantizations of general
relativity in either its canonical or covariant form and string 
theory.  
The latter is characterized by an attempt to construct, at a
fundamental level, a unified quantum theory of all interactions
(sometimes called `theory of everything'), from which quantum gravity
can be recovered in a certain limit. A central problem for all
attempts is the current lack of experimental support. The only
exception is an indirect test of linearized quantum gravity: adopting
the inflationary scenario of the early Universe, the power spectrum of
the CMB is proportional to the Planck time squared and needs the
quantization of the metric for its calculation.\footnote{The Planck
  time is
$t_{\rm P} \equiv \frac{l_{\rm P}}{c}=\sqrt{\frac{\hbar G}{c^5}}
\approx 5.39\times 10^{-44}\ {\rm s}$ ($l_{\rm P}$ is the Planck length.)}
 The density
fluctuations in the CMB have been observed and are in accordance with
this prediction. The influence of primordial gravitons has not been
seen yet, but this is in principle possible; its observation would be a clear test of
(linearized) quantum gravity. 

Oskar Klein's talk at Berne was of a more general nature. Like many
other physicists at the time, he was worried about the divergences in
quantum field theory. In his proposed generalization of general
relativity he went beyond Einstein's own attempts (which he didn't
cite) and discussed the five-dimensional theory which today is known
as Kaluza--Klein theory (but he does not cite Kaluza here). For him,
this theory is the most direct generalization of Einstein's theory
including gauge invariance and charge conservation. As a motivation,
Klein directly referred to nuclear and mesonic physics, for which this
theory should be relevant. He attributed a fundamental role to the
five-dimensional Dirac equation in the sense that it is prior to
geometry: the components of the Riemann tensor follow from the
commutator of the covariant derivatives
\be
\Delta_{\mu}\psi\equiv\left(\frac{\partial}{\partial x^{\mu}}-\Gamma_{\mu}\right)\psi,
\ee
where $\Gamma_{\mu}$ denotes the connection, and $\psi$ is the Dirac
spinor. Gravity is supposed to play an important role when kinetic
energies approach the Planck scale, and Klein speculated that gravity
may serve as a natural regulator for the field theoretical
divergences. In this, he directly related the compactification radius
of the five-dimensional theory to the Planck length.   

Higher dimensions play an important role in string theory, which is
probably consistent only in ten spacetime dimensions. The theory of
supergravity is consistent in any dimension up to eleven; supergravity
may play an important role in the speculative M-theory. 
In string or in supergravity theory, as well as in direct
quantizations of general relativity,  
gravity may directly or indirectly serve as a regulator for the field
theoretic divergences, although the final word has not been spoken
yet.

\section{Summary}

One can state that the Berne Conference of 1955 marks a
turning point in the history of relativity.
This fact is also emphasized in Blum {\em et al.} (2015). It was the first truly
international conference on general relativity and its
generalizations. The importance and prospects of those theories for the future 
are reflected in many contributions to the Proceedings. Today, 
Einstein's theory is empirically well tested, and the fields of
cosmology and quantum gravity occupy a central place in current research.


\section*{Acknowledgements}

I am grateful to the organizers of the conference
{\em Thinking about Space and Time} for inviting me to such an 
inspiring event. I thank Cormac O'Raifeartaigh and Norbert Straumann
for their comments on my 
manuscript and Stanley Deser for sharing his memories of the Berne
Conference.




\begin{thebibliography}{99}

\bibitem{} Ashtekar, A., Bonga, B., and Kesavan, A., 2015.
Asymptotics with a positive cosmological constant:
I. Basic framework. {\em Classical and Quantum Gravity} {\bf 32},
025004 (41pp). 

\bibitem{} Ashtekar, A. and Petkov, V. (eds.), 2014. {\em Springer
    Handbook of Spacetime}. Dordrecht: Springer.

\bibitem{} Blum, A., Lalli, R., and Renn, J., 2015.
  The Reinvention of General Relativity: A Historiographical Framework
  for Assessing One Hundred Years of Curved Space-time.
  {\em Isis} {\bf 106}, 598--620.

 \bibitem{} Bruns, D. G. (2018). Gravitational starlight deflection
measurements during the 21 August 2017
total solar eclipse. {\em Classical and Quantum Gravity} {\bf 35},
075009 (21pp).  

\bibitem{} Chru\'{s}ciel, P. T. and Friedrich, H. (eds.), 2004. {\em
    The Einstein Equations and the Large Scale Behavior of
    Gravitational Fields} (Birkh\"auser, Basel).

\bibitem{} DeWitt, C. (ed.), 1957. {\em Proceedings of the conference
on the role of gravitation in physics}, University of North Carolina,
Chapel Hill, January 18--23, 1957. WADC Technical Report 57-216
(unpublished).
These Proceedings have recently been edited in: D.~Rickles and
C.~M.~DeWitt (eds), Edition Open Sources,
http://www.edition-open-sources.org/sources/5/

\bibitem{} Eckart, A. {\em et al.}, 2017.
  The Milky Way's Supermassive Black Hole: How Good a Case Is It?
  {\em Foundations of Physics} {\bf 47}, 553--624.

\bibitem{} Feynman, R. P., Morinigo, and F. B., Wagner, W. G., 1995. 
{\em Feynman Lectures on Gravitation}. Reading: Addison-Wesley.

\bibitem{} F\"olsing, A., 1994. {\em Albert Einstein}. 3rd ed. Frankfurt am
  Main: Suhrkamp.

\bibitem{} Fuji, Y. and Maeda, K, 2003. {\em The scalar-tensor theory
    of gravitation}. Cambridge: Cambridge University Press.

\bibitem{} Harvey, A. (ed.), 1999. {\em On Einstein's Path}. New York: Springer.

\bibitem{} Isenberg, I., 2014. In: Ashtekar and Petkov (2014),
  pp.~303--321. 

\bibitem{} Kiefer, C., 2012. {\em Quantum Gravity}. 3rd ed. Oxford: Oxford
              University Press.

\bibitem{} Mercier, A. and Kervaire, M. (eds.), 1956. {\em Helvetica
    Physica Acta Supplementum IV} (Birkh\"auser, Basel). This volume
  is available online on https://www.e-periodica.ch.
  
\bibitem{} Planck Collaboration, 2018. Planck 2018
  results. VI. Cosmological parameters. arXiv:1807.06209
  [astro-ph.CO].

\bibitem{} Riess, A. G. {\em et al.}, 2018.
Milky Way Cepheid Standards for Measuring Cosmic Distances and
Application to Gaia DR2: Implications for the Hubble Constant.
{\em Astrophysical Journal} {\bf 861}, 126 (13pp).

\bibitem{} von Meyenn, K. (ed.), 1993. {\em Wolfgang
    Pauli. Wissenschaftlicher Briefwechsel mit Bohr, Einstein,
    Heisenberg u.a.}. Band~III. Berlin: Springer.

\bibitem{} von Meyenn, K. (ed.), 2011. {\em Eine Entdeckung von ganz
    au\ss erordentlicher Tragweite}. {\em Schr\"odingers Briefwechsel
    zur Wellenmechanik und zum Katzenparadoxon}. Berlin: Springer.

\bibitem{} Will, C.~M., 2014. The Confrontation between General
  Relativity and Experiment.  {\em Living Reviews in Relativity},
  available online as  DOI: https://doi.org/10.12942/lrr-2014-4.

\end{thebibliography}
\end{document}